\documentclass{edm_article}
\usepackage{multirow}
\usepackage{array}
\usepackage{layout}
\usepackage{makecell}
\usepackage{dblfloatfix}
\usepackage{booktabs}
\begin{document}

\title{A Comparative Analysis of Student Performance Predictions in Online Courses using Heterogeneous Knowledge Graphs}

\numberofauthors{5}
\author{
Thomas Trask\\
       \affaddr{Georgia Institute of Technology}\\
       \email{thomas.trask@gatech.edu}
\and       
Dr. Nicholas Lytle\\
       \affaddr{Georgia Institute of Technology}\\
       \email{nlytle3@gatech.edu}
\and
Michael Boyle\\
       \affaddr{Georgia Institute of Technology}\\
       \email{mboyle35@gatech.edu}
\and
Dr. David Joyner\\
       \affaddr{Georgia Institute of Technology}\\
       \email{david.joyner@gatech.edu}
\and
Dr. Ahmed Mubarak\\
       \affaddr{IBB University, Yemen}\\
       \email{ahmedmubarak@ibbuniv.edu.ye}
}
\maketitle

\begin{abstract}
As online courses become the norm in the higher-education landscape, investigations into student performance between students who take online vs on-campus versions of classes become necessary. While attention has been given to looking at differences in learning outcomes through comparisons of students' end performance, less attention has been given in comparing students' engagement patterns between different modalities. In this study, we analyze a heterogeneous knowledge graph consisting of students, course videos, formative assessments and their interactions to predict student performance via a Graph Convolutional Network (GCN).  Using students’ performance on the assessments, we attempt to determine a useful model for identifying at-risk students.  We then compare the models generated between 5 on-campus and 2 fully-online MOOC-style instances of the same course. The model developed achieved a 70-90\% accuracy of predicting whether a student would pass a particular problem set based on content consumed, course instance, and modality. 
\end{abstract}

\section{Introduction}
The COVID-19 pandemic accelerated the adoption of online education. In addition to MOOCs and other online-only courses, traditional "on campus" classes made greater use of online course delivery, live or asynchronous discussion, and assessment. However, unresolved concerns about the effectiveness of online courses remain, in particular issues with MOOC-style courses such as low completion rate or low rates of learner engagement \cite{xiong2015examining,kizilcec2015attrition}.  In response, there is an increased use of Learning Analytics to analyze the large volume of data generated by online learning platforms to develop metrics and models to improve learning outcomes.  Among other applications, Learning Analytics has been used to predict learner performance \cite{gardner2018student}, to identify at-risk students in need of early intervention\cite{borrella2022taking, bote2017predicting}, and to drive improvements in the design of online courses \cite{liu2017investigating, outhwaite2020new}.  

In this paper we apply Learning Analytics to Georgia Tech's GTX1301: Introduction to Python course, which offers both "on campus" and "online" versions of the course. In both variants, the same lectures and assessments and other course content are delivered on the EdX platform. However, the learner populations and contexts between the two variants differ, most importantly, the on-campus course is taken by Georgia Tech's undergraduates for a letter grade on a traditional semester calendar, while the online variant is open to any learner without a fixed start or end date. This novel structure allows for opportunities to understand the strengths and limitations of online course delivery and contrast its outcomes with those of a more traditional on-campus learning experience.  Our investigation uses a clickstream dataset provided by the EdX platform, covering five on-campus and two online MOOC-style instances of the course, offered in 2021 and 2022. This interaction data, along with information about the overall course design, was used to construct a Heterogeneous Knowledge Graph (HKG), which serves as input to a Graph Convolutional Network (GCN). This GCN model was then used to predict a learner's likelihood of achieving a passing grade on a given problem set based on the content the learner has previously consumed. A secondary focus of this work was to understand how the models differ between the two course modalities.

\begin{table}[h]
\caption{Course Structure}

\begin{center}
   
\begin{tabular}{|l|r|r|r|r|}
\hline
\textbf{Subcourse}                & \textbf{Videos} & \textbf{\thead{Ungraded \\ exercises}} & \textbf{\thead{Coding \\exercises}} & \textbf{\thead{Graded\\problems}} \\
\hline
\textbf{\thead{Fundamentals}}          & 160             & 56                          & 54                        & 67                       \\
\hline
\textbf{\thead{Control\\structures}}    & 122             & 85                          & 77                        & 85                       \\
\hline
\textbf{\thead{Data\\structures}}       & 117             & 58                          & 44                        & 111                      \\
\hline
\textbf{\thead{Objects \& \\Algorithms}} & 43              & 17                          & 60                        & 32                       \\
\hline
\textbf{Total}                    & \textbf{442}    & \textbf{216}                & \textbf{235}              & \textbf{295}     

 \\ \hline\end{tabular}
\end{center}
\end{table}

\section{Literature Review}

\subsection{In-person vs. Online Classes}
While there had been dedicated work in evaluating and designing online educational opportunities in higher education for decades \cite{castro2021literature}, the Covid-19 pandemic and the rapid switch to remote and then hybrid learning environments accelerated research interest in this space. It became imperative to understand the differences in educational outcomes for students who participate in different modalities of the same course to identify if efforts were needed to close ‘educational gaps’ caused by the pandemic. Fortunately, the switch to remote settings offered instructors an opportune natural experiment to conduct research on these new environments. 
Research comparing course structures of in-person and online has come to several conclusions. First, each modality brings with it a set of affordances that students on the whole respond positively to. Students in online courses enjoy the freedom of having material available on the onset and can better self-regulate their learning by choosing topics, activities, and assessments at their own pace. Affordances of in-person classes tend to center around the social dimensions of courses, especially the face to face and spontaneous interactions between students and instructors as well as among students. Recent research in the structure of online courses has focused on how to replicate these types of interaction for online courses in order to create a better sense of learning community \cite{wang2020jill,kutnick2019synchronous} . 
Traditionally, online courses referred to massively online open courses (MOOCs) where anyone could participate and completion was historically low. However, new online course programs that structure their offerings around the semester system and have a steeper ‘buy-in’ (i.e. enrollment in a degree program, more rigid structure for course activities) has allowed course completion and student achievement to more closely match those found in a traditional degree program \cite{joyner2019replicating,ou2023seven}. These successes point to a trend of blending the affordances of both modalities into offerings that give students the best of all worlds. 
\subsection{Clickstream Data and Neural Networks}
Educational clickstream data has a number of uses in the analysis of learning behavior, from seeing how and when students access or complete resources to redesigning courses to better account for how students engage with material \cite{baker2020benefits}. Past studies have applied a variety of models to clickstream data to predict at-risk learners in self-paced online courses. Al-azazi and Ghurab\cite{al2023ann} used a "day-wise" model with an Artificial Neural Network (ANN) with Long Short Term Memory and predicted student performance with 70\% accuracy in the third month of the course. Rohani\cite{rohani2023early} used an ANN to predict student performance in a MOOC of 3,000 students, achieving an AUC of 81\% with data from the first week of the course. The input to this model was a transition probability matrix rather than an HKG. 
\begin{figure}[b]
    \centering
    \includegraphics[width=1\linewidth]{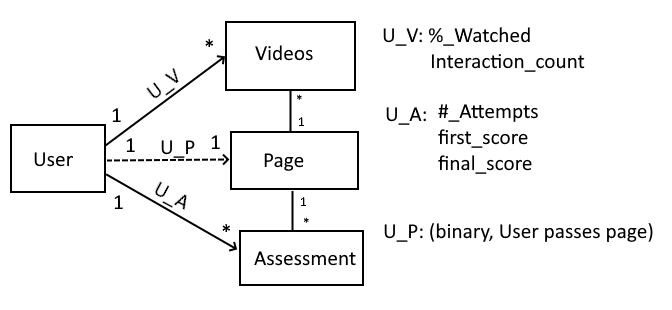}
    \caption{the relationship of users, pages, videos, and assessments.}
    \label{fig:contextoverall}
\end{figure}

Representing the learning environments of online courses as graphs has a number of affordances. Graphs are useful in modeling a student's progression in engaging in learning material at the level of both an individual problem or an entire course \cite{susanti2023link,mubarak2022modeling}. Graph structures as a useful model for online interaction is not unique to educational settings and this usefulness of modeling online structures as a set of nodes and edges has led to the proliferation of graph neural networks. Graph Convolutional Neural Networks have recently emerged as a means of classifying in semi-supervised environments and have been shown to scale better in a wide variety of structured graphs than similar RNN or CNN models alone\cite{kipf2016semi}. 
In prior work \cite{mubarak2022modeling} a Graph Convolutional Network was developed as a semi-supervised classification task to classify students’ engagement in various behavioral patterns. This work first developed a label function to label datasets instead of manual labeling, in which input and output data are labeled for classification to provide a learning foundation for future data processing. Accordingly, four behavioral patterns were hypothesized, namely (“High-engagement”, “Normal-engagement”, “At-risk”, and “Potential-At-risk”) based on students' engagement with course videos and their performance on the assessments/quizzes conducted after. A heterogeneous knowledge graph representing learners was then built using course videos as entities, and capturing semantic relationships among students according to shared knowledge concepts in videos. The model intrinsically works for heterogeneous knowledge graphs as a semi-supervised node classification task. It was evaluated on a real-world dataset across multiple settings to achieve a better predictive classification model. While this work was successful in classifying students in a real-world setting, this current context extends prior work in two ways. First, our course context with both on-campus and on-line learners allows us to compare student engagement behavior depending on modality of learning. Second, this work includes assessment data, providing ground truth values in student scores on various assessments.
\section{Methodology}
\subsection{Course Context}
Our data comes from several semesters of  Georgia Tech's GTX1301: Introduction to Python course, which offers both "on campus" and "online" versions of the course. The course matches a standard introduction to computer science course covering topics such as basic program construction. The course is intended for students with zero or near-zero programming experience. In both variants, the same lectures and assessments are delivered on the EdX platform, which has typically been used for online-only MOOC courses. Table 1 shows the number of students who engaged and completed all course requirements for each semester and modality. The course content and instructional staff was the same for all offerings.  
Table 1 gives an overview of the course structure for the online course instances.  The online instances are divided into 4 submodules and users could choose to take any of the four submodules.  Each submodules is composed of 3-7 chapters, delivering course content primarily through short videos and texts. Formative assessments inclduded multiple-choice exercises, short answer exercises, and interactive coding exercises which were checked but did not contribute to the learner's final grade. There were graded problem sets at the end of each unit, as well as a graded final test (and ungraded practice tests) at the end of each module. Online learners had to pay a fee to have access to the graded problem sets and the graded final test.

The original dataset consists of ~16M clickstream events from the 5 on-campus course instances and ~55M events from the 2 online course instances, each representing a year's worth of user interaction. This data was then filtered to focus on user-initiated clickstream events, namely video interactions and assessment submissions as these categories of event were more directly relevant to understanding how users interacted with the course content. Clickstream events originating from the edX mobile app were also removed, as there was not a comparably straightforward way to link page content together for those records. The data was then collated into the following datasets:
\begin{enumerate}
    \item Student feature matrix which includes the following student metadata: session count, total counts for various browser interactions, first, last access date, and student interaction level based on the calculation found in appendix B. Each student was labeled with one of the following interaction levels: high engagement, normal engagement, potential at-risk, or at-risk.
    \item Video/assessment feature matrices consisting of video duration.
    \item Course Content Edge Matrix:  Each video and assessment link to a page.  Videos and assessments were grouped together by whether they shared the same [page] property.  
    \item Student->Content edge matrix:  Each student-video edge weight was calculated based on the final \% of the video the student watched, normalized. Each student-assessment edge represents a feature set containing the users first grade, final grade, and number of attempts. 
    \item Student->Page Edge Matrix:  A student->page edge is a binary representation of whether a student met the page success criteria .  For pages containing assessments, this represents whether the student achieved an average score of >70\% on all page assessments.   For pages containing only videos, this represents whether the user had watched >70\% of the page's videos.
\end{enumerate} 

\begin{table}
\caption{Course User/Interaction Totals}

\begin{tabular}{|l|l|r|r|r|}

\hline
\textbf{Modality}                   & \textbf{Term}  & \textbf{Users} & \textbf{\thead{Browser \\Events}} & \textbf{Sessions} \\
\hline
\multirow{6}{*}{\textbf{On campus}} & Spring 2021    & 341            & {113,016}                 & 2,120             \\
                                    & Summer 2021    & 94             & {27,849}                  & 276               \\
                                    & Fall 2021      & 513            & {1,607,457}               & 4,011             \\
                                    & Spring 2022    & 371            & {1,179,669}               & 1,386             \\
                                    & Summer 2022    & 288            & {849,683}                 & 1,942             \\
                                    & \textbf{Total} & \textbf{1,607} & {\textbf{3,777,675}}      & \textbf{9,735}    \\
\hline
\multirow{3}{*}{\textbf{Online}} & 2021        & 25,670         & {11,833,984}                                                      & 79,664            \\                                                                       
                                    & 2022       & 21,346          & {3,916,560}                 & 24,386
             \\
                                    & \textbf{Total} & \textbf{47,016} & {\textbf{15,750,544}}     & \textbf{104,050} 

\\
\hline\end{tabular} 

\end{table}

\begin{figure*}[h]
    \centering 
    \includegraphics[width=1\linewidth]{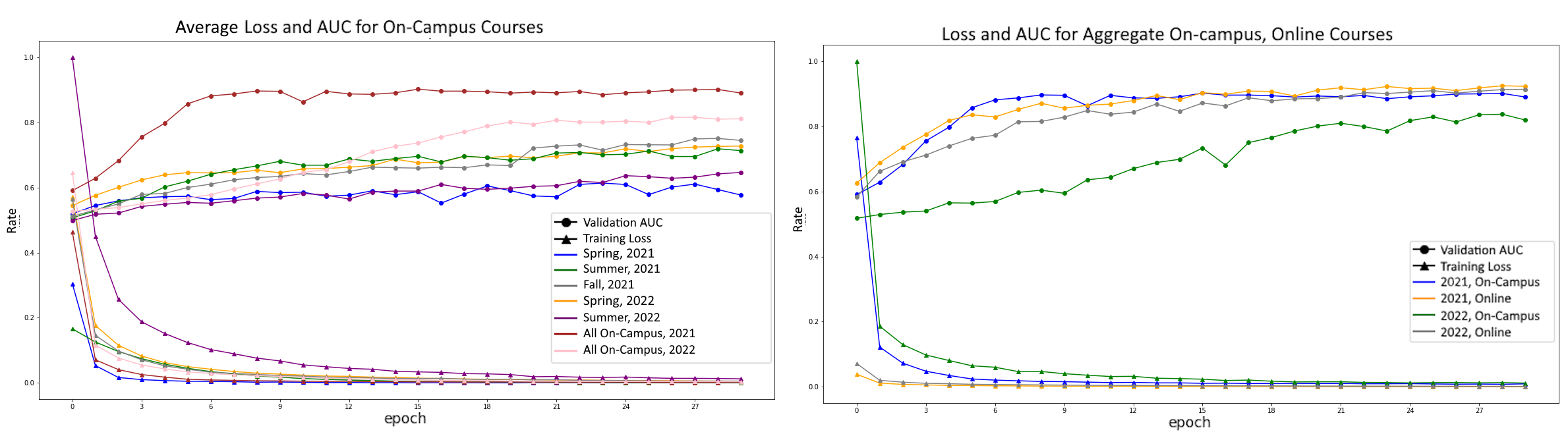}
    \caption{GCN Model Loss/AUC Results, On-campus vs Online}
    \label{fig:contextoverall}
\end{figure*}

\subsection{Data Limitations}
As seen in Figure 1, course videos and assessments are linked together via course “pages”.  Although there is a ‘page’ field in each EdX log record,  only ~30\% of course videos and assessments could be directly linked to course pages via that property.   Only assessments records had associated display name within the logs; course videos did not.  More research must be done to determine a more robust method of linking course content. 

\subsection{Data Governance}
As noted, the clicktream data used in this experiment was gathered from multiple instances of the GTX1301 on the edX platform.  As part of edx's platform, users agree to automated collection of the tracking data in the experiment. Data was gathered automatically while users interacted with the edx course site in accordance with the GDPR and openEdx data governance guidelines.\cite{edxTos}  Each data record included a user's edx user ID, along with the datetime of the record's event. Other than two polls indicating a user's previous experience with course material, no demographic information was included in the dataset.  Data governance for the experience was managed via GT IRB.  During the experiment, the dataset was accessed by GT researchers through GT private cloud and a secured, offline dataset instance.  

\subsection{Graph Convolution Model}
This heterogeneous knowledge graph structure above was converted into a Pytorch Geometric (PyG) graph convolutional network (GCN) for analysis.  The GCN uses a 3 stage classifier with 2 Sage Convolution layers and a rectified linear unit (ReLU) activation function between them.  Input node data was processed using a standard embedding and linear transform combination..  The dot product between the user and page nodes was then taken to derive edge-level predictions of whether a student 'passed' a particular page.  The GCN was trained using 64 hidden layers and 4 output layers.  For training the model, the GCN was split into an 80/10/10 subgraph training split based on users’ interaction with course content using the PyG Random Link Splitter, similar to how standard subsampling splits a dataset along rows.  Across 30 epochs, the training data was split into subgraphs using the PyG link neighbor loader, and trained using the PyTorch binary cross entropy with logits loss function with the Adam optimizer to calculate total loss, in-situ AUC, and final AUC validation.  See Appendix A for more information regarding the GCN.

\section{Results and Discussion}
The GCN model achieved an average AUC score of ~58-90\% predicting whether a student will pass a particular problem check.  Fig 2 shows the model's loss and AUC for each individual on-campus course instances, both online course instances, and an aggregate of each year's on-campus courses, averaged across 5 iterations.  

For  on-campus course instances, the AUC the GCN model varied widely, both comparatively between courses and between individual runs for courses.  Fig. 2 shows average AUC scores of individual courses ranging from 50-80\%.  It shows that the GCN predictor performs better when looking at each year's data in aggregate; the AUC score rises to 90\% for 2021 and 82\% for 2022.  The lower AUC for 2022 is likely a result of missing data for the Fall 2022 semester.  This is also evident when comparing the model results for the aggregate on-campus and online programs.  While the 2021 on-campus CGN model performs in lock-step with the 2021 and 2022 online GCN models, the 2022 on-campus model lags.  More research is needed to unpack the variance in results for the on-campus course instances as there is no direct correlation between a courses user count, interaction count, session count, variance, or combinations thereof.  
\begin{figure}[!b]
    \centering
    \includegraphics[width=1\linewidth]{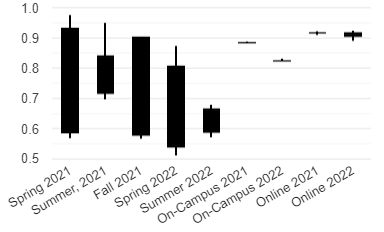}
    \caption{GCN Model AUC Variance, On-Campus vs Online}
    \label{fig:contextoverall}
\end{figure}
\subsection{Repeatability and Transferability}
It is important to note that there is significant variance within and between GCN training runs when looking at any individual course instance, as see in fig. 3.  This is likely due to a floor effect when splitting the data for the individual on-campus courses.  As noted, graph training was done on an 80/10/10 split along individual user interactions.  It is theorized that, as the course user count lowers, course data may be more susceptible to successful/unsuccessful students being grouped together, impacting results. This is indicated in the relatively higher per-run variance in AUC scores for the individual on-campus course instances when compared to the aggregated on-campus courses and online course instances, where the user and interaction counts are much higher.  While the online classes generate an order of magnitude more data, additional analysis needs to be conducted to determine the cause and impact of this difference in variance.  
Even though the GCN model successfully generated prediction models for each course instance, the difference in the data’s shape caused the model for one class to be non-transferable to other course instances.  A translation layer must be developed to normalize available course content between courses to solve this transferability issue.

\section{Conclusion and Limitations}
This work extends previous clickstream GCN models to include student assessment interaction but focused on one course offering. Further investigations into different courses, topics, and degree paths are needed. Our understanding of the demographics of the students who engaged in these course offerings was quite limited. As the project continues, we plan on developing more robust implementations of the Graph network to include these demographics.

In future work, we plan to include more information from student engagement into the model in order to extend predictions. This can include information from participation in class forums like Piazza, better content-mapping between course concepts and content, and more fine-grained video interaction patterns.  Going forward, a scalable method will need to be developed to infer connections between different content for each course.  A  transposition layer should then be created to link course content across courses and transfer one course instance's GCN model to other course instances.   Finally, exploring user interaction via temporal graphs should allow us to examine user interactions on a per-session and per-interaction-group basis to determine whether key interaction patterns for target user groups exist.  
\break
%
\bibliographystyle{abbrv}
\bibliography{sigproc}  
%
\appendix

\section{GCN Platform Metrics} 
The below system was used to aggregate the log data and train the GCN used in this exercise:
\begin{itemize}
    \item CPU: Intel Core i5 12600 10-core 16-thread CPU
\item GPU: Nvidia Geforce RTX 3060ti GPU 
\item Memory: 64GB DDR4 3200mts
\item Disk: 512GB NVME PCIE 3.0 SSD
\item OS: Windows 10 2023H2
\item Language: Python 3.11
\item MongoDB 7.0, hosted locally
\item Software Environment: Spyder IDE 5.4.3
\item GCN Training Hyperparameters:  
\begin{enumerate}
\item epoch count: 30
\item learning rate: 0.001
\item batch size: 64
\item Link Neighbor Loader neighborhood: [4,2]*4    
\end{enumerate}
\end{itemize}
On the above system, data generation and GCN training times scaled semi-linearly based on the size of the course’s content and its’ user count.  Data generation and training for [on-campus Spring 2021, 94 users] took roughly 20 seconds, whereas the Online, 20211, 21k users took 22 minutes.   GCN hyperparameters were chosen based on a combination of computational time efficiency and accuracy.  Modifying the learning rate and Link neighbor loader neighborhood shape had the largest impact on efficiency and accuracy.  The system used the CUDA framework for GCN training, resulting in a 0.5-3x performance gain over GCN training using the CPU.  System performance will vary and can be improved by leveraging the Python multithreading libraries.  Please feel free to reach out to the researchers for copies of the GCN processor and Data aggregator.

\section{Student Interaction Level Calculation}
\begin{figure}[h]
    \centering
    \includegraphics[width=1\linewidth]{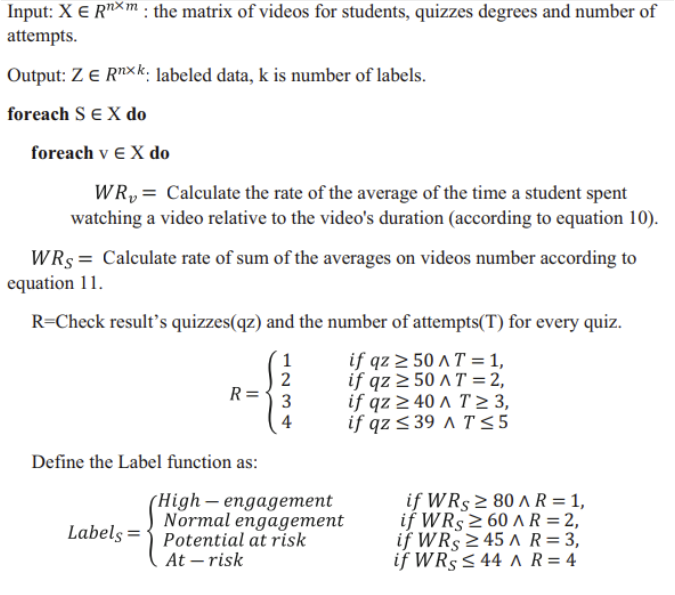}
    \caption{The calculation used to determine a student’s engagement level.\cite{mubarak2022modeling}}
    \label{fig:contextoverall}
\end{figure}

\balancecolumns
\end{document}